\documentclass[twocolumn,showpacs,preprintnumbers,amsmath,amssymb]{revtex4}

\usepackage{graphicx}
\usepackage{dcolumn}
\usepackage{bm}

\begin{document}


\title{An Electromagnetic GL Double Layered Cloak}

\author{Ganquan Xie}
 \altaffiliation[Also at ]{GL Geophysical Laboratory, USA, glhua@glgeo.com}
\author{Jianhua Li, Feng Xie, Lee Xie}%
 \email{GLGANQUAN@GLGEO.COM}
\affiliation{%
GL Geophysical Laboratory, USA
}%

\hfill\break

\date{\today}

\begin{abstract}
In this paper, we propose a new electromagnetic (EM) GL double layered cloak.
The GL double layered cloak is consist of two sphere annular layers,
$R_1 \le r \le R_2$ and $R_2 \le  r \le R_3$. Two type cloak materials
are proposed and installed in the each layer, respectively.
The outer layer cloak of the GL double layered cloak has the invisible function,
the inner layer cloak has fully absorption function. The GL double layered 
metamaterials are weak degenerative and weak dispersive.
When the source is located outside 
of the GL double layered cloak, the excited EM wave field 
propagation outside of the double layered cloak is as same as in free space and never be disturbed 
by the cloak; also, the exterior EM wave can not penetrate into the
inner layer and concealment.
When local sources are located inside of the GL double cloaked concealment
with the normal EM materials, the 
excited EM wave is propagating under Maxwell equation governing,
it is complete absorbed by the inner layer cloak of GL double cloak 
and never propagate to outside of the inner layer of the GL cloak,
moreover, the EM wavefield in concealment never be disturbed by the cloak. 
The GL doubled layered cloak is a robust cloak and has complete and sufficient
invisibility functions. Its concealment is the normal electromagnetic
environment. Our EM GL double layered cloak is different from
conventional common cloak.
The 3D GL EM modeling simulations for the double layered cloak
are presented. The GL method is an effective physical simulation method and is fully 
different from the conventional methods. It has 
double abilities of the theoretical analysis and numerical simulations to study the 
cloak metamaterials and wide materials and field scattering problem in physical sciences.
\end{abstract}

\maketitle

\section{\label{sec:level1}INTRODUCTION} 
 
Using the 3D GL EM modeling [1-2] and GL inversion [3], we propose an EM double layered cloak in this paper which is called as
GL double layered cloak.
The single layer cloak proposed by 
Pendry et al. [4] is named as PS cloak.  The GL double layered cloak is consist of two sphere annular layers,
$R_1 \le r \le R_2$ and $R_2 \le  r \le R_3$. Two different type cloak materials
are proposed and installed in the each layer, respectively.
The outer layer cloak of the GL double layered cloak has the invisibility function,
while the inner layer cloak has fully absorption function to absorb
the EM wave excited from local sources inside of the 
concealment.
When the  source is located outside 
of the GL double layered cloak, the excited EM wave field 
propagation outside of the double layered cloak is as same as in free space and never be disturbed 
by the cloak; also, the exterior EM wave can not penetrate into the
inner layer and concealment.
When local sources are located inside of the GL double cloaked concealment
with the normal EM materials, the 
excited EM wave is propagating under Maxwell equation governing,
it is complete absorbed by the inner layer cloak of GL double cloak 
and never propagate to outside of the inner layer of the GL cloak,
moreover, the EM wavefield in concealment never be disturbed by the cloak. 

The GL doubled layered cloak is a robust cloak and has complete and sufficient
invisibility functions. Its concealment is the normal electromagnetic
environment. Our EM GL double layered cloak is different from
conventional common cloak. The GL double layered 
metamaterials are weak degenerative and dispersive.

A double layer cloth phenomenon to prevent the GILD inversion [5-7]
detection has been observed in paper [8] in 2001 which is published in SEG online
http://www.segdl.org/journals/doc/SEGLIB-home/dci/searchDCI.jsp. 
We developed a novel and effective Global and Local field (GL) modeling and inversion to
study the metamaterials, periodic photonic crystals and condense
physics etc. wide physical sciences. 
3D GL EM modeling and inversion [9] and computational mirage [10] have been presented in PIERS 2005 
and published in
proceeding of PIERS 2005 in Hangzhou, which can be downloaded from 
http://piers.mit.edu/piersproceedings/piers2k5Proc.php.
The analytical method and numerical method for physical
sciences are developed separately in history. The GL method consistent combines the analytical
and numerical methods together. The GL method does not need to solve large matrix equation,
it only needs to solve $3 \times 3$ and $6 \times 6$  matrix equations. Moreover, the GL method
does not need any artificial boundary and not need PML absorption condition on it to
truncate the infinite domain. 
The Finite Element method (FEM) and Finite Difference (FD)
method have numerical dispersions which confuse and
contaminate the physical dispersion from the interaction
between the field and materials. The frequency limitation
is a difficulty of FEM and FD method.

The GL method is a significant scattering process which reduces the numerical dispersion and
is suitable to simulate physical wavefield scattering in the 
materials, in particular, for dispersive materials.
Born Approximation is a conventional method in the quantum mechanics
and solid physics, however, it is one iteration only in whole domain which is
not accurate in the high frequency and high contrast materials. The GL method divides the domain as a set of small sub domains or sub lattices. The Global field is updated by the local field 
from the interaction between the global field  
and local subdomain materials successively. Once all subdomain
materials are scattered, the GL field solution is obtained which
is much more accurate than the Born approximation.
 
Moreover, the GL method can be meshless, including arbitrary geometry
subdomains, such as rectangle, cylindrical and spherical coordinate
mixed coupled together. It is full parallel algorithm.  
These advantages of the GL method to overcome historical
difficulties have been detailed described in the paper [1]. 
The theoretical foundation of the GL method is described in the paper [2]

We have used the 3D GL modeling [1-2] and inversion [3] to simulate 
many cloak metmaterials, nanometer materials, periodic photonic crystals etc.
When the point source is located outside or inside of the various geometry cloaks, 
the 3D GL EM modeling simulations for the EM wavefield propagation through the cloaks
have been done. These simulations show that the GL method is fast and accurate.
We have submitted a paper titled the 3D GL EM modeling to simulate single layer 
cloaks to PRE [11]. 

In this paper, the 3D GL EM modeling simulations of the EM wave field propagation through
the new GL double layered cloak is presented. When the local sources are located outside or
inside of the outer layer cloak, EM wave propagation through outer layer cloak 
and never penetrate into the inner layer and the concealment, i.e. $r \le R_2$. 
The exterior EM wavefield propagation outside of GL double layered cloak never
be disturbed by the cloak. The outer 
layer cloak has the invisibility  function. When the local sources 
inside of the GL double layer cloaked concealment with normal materials, the excited EM wave normally
propagating under Maxwell equation governing, the EM field is complete absorbed
by the inner layer cloak and can not propagate outside of the inner cloak.
The EM environment in the GL double cloaked concealment is normal, in which
there exist Maxwell EM wave field excited by nonzero local sources, have no reflection 
from the boundary $r=R_1$, and never propagate outside of boundary $r=R_2$  

By using the 3D GL EM modeling [1-2] simulation  and its theoretical analysis, 
we found and verified a phenomenon that there exists no Maxwell EM wave field
can be excited by nonzero local sources inside of the single layer cloaked concealment
with normal materials.
Our GL double layered cloak overcomes the drawback and difficulty in the single
cloak. Pendry et al. in paper \cite{4} used a coordinate transformation and ray 
tracing 
to propose the annular cloak in which the ray
being bending and re direction around central sphere object and can not 
penetrate into it.
The cloak device like empty and does not disturb exterior wave field.
There are several other papers to simulate 
the exterior plane wave propagation through the cloak [12-14]. 
Cummer et al. in paper \cite{12} proposed numerical simulations by using
the COMSOL Multiphysics finite element-based electromagnetics
solver for the 
2D plane wave propagation through cylindrical cloak. Chen et al. proposed the 
Mie analytical TEM model to simulate the plane 
wave through the spherical cloak \cite{13}. 
Argyropoulos et al. 
proposed a dispersive finite difference method in time 
domain (FTFD) in [14] to simulate 2D TEM plane wave field through cylindrical cloak, in which 
authors considered the difficulty of conventional FDTD scheme 
for dispersive materials. In papers \cite{12} and \cite{14}, authors 
introduced many papers for cloak research works.
Because the plane wave is excited by plane source 
which can not be located inside of the cloak or concealment. 
To study the EM wave excited from local sources inside 
of the cloaked concealment is absent from these papers.
In paper [15] and [16], authors studied the effect on invisibility of active devices 
inside the cloaked region. Author in [16] stated that
"when these conditions are overdermined, finite energy solutions
typically do not exist."

We use 3D GL method to do many simulations for
studying the behavior of EM field excited inside of the single
layer cloaked concealment. These simulation are divergent or
become chaos when the EM wave propagates to the inner
boundary of the single layer cloak.  
Our statement is that "there exists no Maxwell EM wave field
can be excited by nonzero local sources inside of the single layer cloaked concealment
with normal materials".
The detailed proof and 3D GL simulations are presented in this paper. 
Before the practice production of the single layered cloak,
the electromagnetic field environment inside of the concealment 
can not be studied in physical experiment.
Our GL double layered cloak proposed in this paper
overcomes the drawback and difficulty of the
single layer cloak and avoid the disputing on the 
EM phenomenon inside of the single layered cloaked concealment
with normal materials. The GL double layered EM cloak 
metamaterials inventive and fabrication technology right
and 3D GL EM modeling software are patented by GL Geophysical Laboratory.
We thanks to GL Geophysical Laboratory to approve us to publish
the GL modeling method, GL double layered cloak theory, and simulations.

We describe this paper in the following order: The introduction is described in 
Section 1. In Section 2, we propose a GL double layered
cloak materials. The EM integral 
equations are presented in Section 3. The 3D GL EM modeling are described 
in Section 4. The theoretical analysis of properties and functions of the GL double
layered cloak are proposed in Section 5.
The simulations of the EM wave propagation through the GL double layered Cloak 
by using the GL EM modeling are presented in Section 6.
The advantages of the GL double layered cloak is presented in Section 7.
In Section 8, we conclude this paper.

\section{\label{sec:level1} GL Double Layered Cloak Materials}

\subsection {GL Inner Layered Cloak Anisotropic Material}
On the inner sphere annular domain, $\Omega _{GLI}  = \left\{ {r:R_1  \le r \le R_2 } \right\},$  by the GL EM modeling and inversion [1-3],
we propose an anisotropic material as follows,
\begin{equation}
\begin{array}{l}
 \left[ D \right]_{GLI}  = diag\left[ {\bar \varepsilon _i ,\bar \mu _i } \right], \\ 
 \bar \varepsilon _i  = diag\left[ {\varepsilon _{r,i} ,\varepsilon _{\theta ,i} ,\varepsilon {}_{\phi ,i}} \right]\varepsilon _0,  \\ 
 \bar \mu _i  = diag\left[ {\mu _{r,i} ,\mu _{\theta ,i} ,\mu {}_{\phi ,i}} \right]\mu _0,  \\ 
 \varepsilon _{r,i}  = \mu _{r,i}  = \left( {\frac{{R_2^2  - R_1^2 }}{{R_2^2 }}} \right)\sqrt {\frac{{R_{^2 }^2  - r^2 }}{{R_2^2  - R_1^2 }}},  \\ 
 \varepsilon _{\theta ,i}  = \varepsilon _{\phi ,i}  = \mu _{\theta ,i}  = \mu _{\phi ,i}  = \sqrt {\frac{{R_2^2  - R_1^2 }}{{R_{^2 }^2  - r^2 }}} \frac{{R_2^2 }}{{R_{^2 }^2  - r^2 }}. \\ 
 \end{array}
\end{equation}
The $\Omega_{GLI}$ is called as GL inner layered cloak, the materials, $ \left[ D \right]_{GLI}  = diag\left[ {\bar \varepsilon _i ,\bar \mu _i } 
\right]$ in (1), are the anisotropic GL inner layered cloak material
tensor.

\subsection {GL Outer Layered Cloak Anisotropic Material}
Let the outer sphere annular domain $\Omega _{GLO}  = \left\{ {r:R_2  \le r \le R_3 } \right\}$ be the GL outer layered cloak
with the following anisotropic GL outer layered cloak materials,
\begin{equation}
\begin{array}{l}
 \left[ D \right]_{GLO}  = diag\left[ {\bar \varepsilon _o ,\bar \mu _o } \right], \\ 
\bar \varepsilon _o  = diag\left[ {\varepsilon _{r,o} ,\varepsilon _{\theta ,o} ,\varepsilon {}_{\phi ,o}} \right]\varepsilon _0, \\ 
 \bar \mu _o  = diag\left[ {\mu _{r,o} ,\mu _{\theta ,o} ,\mu {}_{\phi ,o}} \right]\mu _0,  \\ 
  \varepsilon _{r,o}  = \mu _{r,o}  = \frac{{R_3 }}{r}\frac{{r^2  - R_2^2 }}{{r^2 }}\frac{{\sqrt {r^2  - R_2^2 } }}{{\sqrt {R_3^2  - R_2^2 } }}, \\ 
 \varepsilon _{\theta ,o}  = \mu _{\theta ,o}  = \varepsilon _{\phi ,o}  = \mu _{\phi ,o}  \\ 
  = \frac{{R_3 }}{{\sqrt {R_3^2  - R_2^2 } }}\frac{r}{{\sqrt {r^2  - R_2^2 } }}. \\ 
\end{array}
\end{equation}

The GL inner cloak $\Omega_{GLI}$ domain and GL outer cloak $\Omega_{GLO}$ domain 
are bordering on the interface surface $r=R_2$.
We assemble the $\Omega_{GLI}$ as the inner layer sphere annular domain and $\Omega_{GLO}$ as
the outer layer sphere annular domain and make them coupling on their interface boundary annular surface  $r=R_2$
as follows,
\begin{equation}
\begin{array}{l}
 \Omega _{GL}  = \Omega _{GLI} \bigcup {\Omega _{GLO} }  \\ 
  = \left\{ {r:R_1  \le r \le R_2 } \right\}\bigcup {\left\{ {r:R_2  \le r \le R_3 } \right\}}  \\ 
  = \left\{ {r:R_1  \le r \le R_3 } \right\}, \\ 
 \end{array}
\end{equation}
and offer the coupled anisotropic dielectric and susceptibility tensor $\left[ D \right]_{GL}$ on the $\Omega _{GL}$  as follows,
\begin{equation}
\left[ D \right]_{GL}  = \left\{ {\begin{array}{*{20}c}
   {\left[ D \right]_{GLI} ,r \in \Omega _{GLI} }  \\
   {\left[ D \right]_{GLO} ,r \in \Omega _{GLO} }.  \\
\end{array}} \right.
\end{equation}
The GL inner layer cloak material $ \left[ D \right]_{GLI}  = diag\left[ {\bar \varepsilon _i,\bar \mu _i } 
\right]$ in (1) on the $\Omega _{GLI}$  and GL outer layer cloak material $ \left[ D \right]_{GLO}  = diag\left[ {\bar \varepsilon _o ,\bar \mu _o } 
\right]$ in (2) on $\Omega _{GLO}$  are assembled into the GL anisotropic double layered cloak material on
the domain $\Omega _{GL}$. The domain $\Omega _{GL}$ with the metamaterial $\left[ D \right]_{GL}$ in (4) 
is called as the  GL double layered cloak.

\section{\label{sec:level1}3D ELECTROMAGNETIC INTEGRAL EQUATION}
The 3D EM integral equation in frequency domain has been proposed in authors' papers
[1] and [2].  In this  paper,  we proposed the EM integral equation in time domain as follows:

\begin{equation}
\begin{array}{l}
 \left[ {\begin{array}{*{20}c}
   {E(r,t)}  \\
   {H(r,t)}  \\
\end{array}} \right] = \left[ {\begin{array}{*{20}c}
   {E_b (r,t)}  \\
   {H_b (r,t)}  \\
\end{array}} \right] \\ 
  + \int\limits_\Omega  {G_{E,H}^{J,M} (r',r,t) * _t \delta \left[ {D(r')} \right]\left[ {\begin{array}{*{20}c}
   {E_b (r',t)}  \\
   {H_b (r',t)}  \\
\end{array}} \right]dr'}, \\ 
 \end{array}
\end{equation}
and
\begin{equation}
\begin{array}{l}
 \left[ {\begin{array}{*{20}c}
   {E(r,t)}  \\
   {H(r,t)}  \\
\end{array}} \right] = \left[ {\begin{array}{*{20}c}
   {E_b (r,t)}  \\
   {H_b (r,t)}  \\
\end{array}} \right] \\ 
  + \int\limits_\Omega  {G_{E,H,b}^{J,M} (r',r,t) * _t \delta \left[ {D(r')} \right]\left[ {\begin{array}{*{20}c}
   {E(r',t)}  \\
   {H(r',t)}  \\
\end{array}} \right]dr'}.  \\ 
 \end{array}
\end{equation}
In the EM integral equation (5), 
\begin{equation}
G_{E,H}^{J,M} \left( {r',r,t} \right) = \left[ {\begin{array}{*{20}c}
   {E^J \left( {r',r,t} \right)} & {H^J \left( {r',r,t} \right)}  \\
   {E^M \left( {r',r,t} \right)} & {H^M \left( {r',r,t} \right)}  \\
\end{array}} \right],
\end{equation}
and $G_{E,H,b}^{J,M}$ is the EM Green's tensor in the background medium,
where, ${E(r,t)}$ is the electric field, ${H(r,t)}$ is the magnetic field, $E_b (r,t)$ and
 $H_b (r,t)$ is the incident electric and magnetic field
 in the background medium, $E^J (r',r,t)$ is electric Green's tensor, $H^J (r',r,t)$ is 
 magnetic Green's tensor, they are excited by the point impulse current source,
$E^M(r',r,t)$ and $H^M (r',r,t)$ are electric and magnetic Green's tensor, respectively,
 they are excited by the point impulse magnetic moment source,
 $ * _t$ is convolution with respect to t,
$\delta \left[ D \right]$ is the electromagnetic material  parameter variation matrix,
\begin{equation}
\begin{array}{l}
 \delta \left[ D \right] = \left[ {\begin{array}{*{20}c}
   {\delta D_{11} } & 0  \\
   0 & {\delta D_{22} }  \\
\end{array}} \right], \\ 
 \delta D_{11}  = (\bar \sigma (r) - \sigma _b I) + (\bar \varepsilon (r) - \varepsilon _b I)\frac{\partial }{{\partial t}}, \\ 
 \delta D_{22}  = (\bar \mu (r) - \mu _b I)\frac{\partial }{{\partial t}}, \\ 
 \end{array}
\end{equation}
$\delta D_{11} $ and $\delta D_{22}  $ are a $3\times 3$ symmetry, 
inhomogeneous diagonal matrix  for  the isotropic material,
for anisotropic material, they are an inhomogeneous diagonal or full matrix, 
$I$ is a $3\times 3$ unit matrix, $\bar \sigma (r)$ is the conductivity tensor,
$\bar \varepsilon (r)$ is the dielectric tensor, $\bar \mu (r)$ is susceptibility tensor which can be dispersive
parameters depend on the angular frequency $\omega$, 
$\sigma _b$ is the conductivity,  $\varepsilon_b $
is the permittivity, $ \mu _b$ is the permeability in the background free space, $\Omega$
is the finite domain in which the parameter variation matrix $\delta \left[ D \right] \ne 0,$
the $(\bar \varepsilon (r) - \varepsilon _b I)E $ 
is the electric polarization, and $(\bar \mu (r) - \mu _b I)H $ is the magnetization.

\section{\label{sec:level1}3D GL EM MODELING}
We propose the GL EM modeling based on the EM integral equations (1) and (2) in the time space
domain.

 (3.1)	The domain $\Omega$ is divided into a set of $N$ sub domains,$\{\Omega_k\}$, 
such that $\Omega  = \bigcup\limits_{k = 1}^N {\Omega _k }$. The division can be mesh or meshless.

(3.2)  When $k=0$, let
$E_0 (r,t)$ and $H_0 (r,t)$ are the analytical global field, $E_0^J (r',r,t)$, $H_0^J (r',r,t)$, $E_0^M (r',r,t)$, and
$H_0^M (r',r,t)$ are the analytical global Green's tensor in the background medium. By induction, suppose that
$E_{k-1} (r,t)$, $H_{k-1} (r,t)$, $E_{k-1}^J (r',r,t)$, $H_{k-1}^J (r',r,t)$, $E_{k-1}^M (r',r,t)$, and
$H_{k-1}^M (r',r,t)$ are calculated in the $(k-1)^{th}$ step in the subdomain $\Omega_{k-1}$.

(3.3) In $\{\Omega_k\}$, upon substituting $E_{k-1} (r,t)$, $H_{k-1} (r,t)$, $E_{k-1}^J (r',r,t)$, $H_{k-1}^J (r',r,t)$, $E_{k-1}^M (r',r,t)$, and
$H_{k-1}^M (r',r,t)$ into the integral equation (1),  the EM Green's tensor integral equation (1)
in $\Omega_{k}$ is reduced into $6\times 6$ matrix equations. By solving the $6\times 6$  matrix equations, 
we obtain the Green's tensor field  $E_{k}^J (r',r,t)$, $H_{k}^J (r',r,t)$, $E_{k}^M (r',r,t)$, and
$H_{k}^M (r',r,t)$.

(3.4) According to the  integral equation (1), the electromagnetic field $E_{k} (r,t)$ and $H_{k} (r,t)$ are updated by 
the interaction scattering field between the Green's tensor and local polarization and magnetization in the subdomain $\Omega_{k}$ as follows,
\begin{equation}
\begin{array}{l}
 \left[ {\begin{array}{*{20}c}
   {E_k (r,t)}  \\
   {H_k (r,t)}  \\
\end{array}} \right] = \left[ {\begin{array}{*{20}c}
   {E_{k - 1} (r,t)}  \\
   {H_{k - 1} (r,t)}  \\
\end{array}} \right] \\ 
  + \int\limits_{\Omega_k}  {\left\{ {\left[ {\begin{array}{*{20}c}
   {E_k^J (r',r,t)} & {H_k^J (r',r,t)}  \\
   {E_k^M (r',r,t)} & {H_k^M (r',r,t)}  \\
\end{array}} \right]} \right.}  \\ 
 \left. { * _t \delta \left[ {D(r')} \right]\left[ {\begin{array}{*{20}c}
   {E_{k - 1} (r',t)}  \\
   {H_{k - 1} (r',t)}  \\
\end{array}} \right]} \right\}dr' \\ 
 \end{array}
\end{equation}

(3.5) The steps (3.2) and (3.4) form a finite iteration, $k = 1,2, \cdots, N$,
the $E_N \left( r,t \right)$ and $H_N \left( r,t \right)$
are the electromagnetic field  of the GL modeling method. The GL
electromagnetic field modeling in the time space domain is short named as GLT method.

The GL EM modeling in the space frequency domain is proposed in the paper \cite{2}, 
we call the GL modeling in frequency domain as GLF method. 

\section{\label{sec:level1}THEORETICAL ANALYSIS OF INTERACTION OF 
THE EM WAVE FIELD THROUGH THE CLOAKS}

\subsection {Theoretical Analysis Of Interaction Of 
The EM Wave Field Through The GL Double Layered Cloaks}

We propose the theoretical analysis of the interaction between the EM wave and GL cloaks in this section.

$\textbf{Statement 1:}$  Let domain $\Omega _{GL}$ in (3) and the metamaterial $D_{GL}$ in (4) be GL double layered cloak, and
 $\varepsilon {\rm  = }\varepsilon _{\rm b},\mu  = \mu _b$
be basic permittivity and permeability, respectively, 
inside of the central sphere concealment $| \vec {r'}  | < R_1$ and outside of the GL cloak $ | \vec {r'} | >R_3$,  
we have the following statements:
(1) provide the source is located inside 
of the concealment of GL double layered cloak, $|\vec r_s |< R_1$, the excited EM wave field propagates inside of the concealment and
never be disturbed 
by the cloak; (2) provide the source is located inside 
of concealment or inside of the inner layer  of the GL double layered cloak,
$|\vec r_s| < R_2,$ the EM wave field is vanished outside of the inner layer of GL cloak
and is always propagating to the boundary $r=R_2$ and is absorbed by the boundary $r=R_2$.
(3) provide the  source is located outside 
of the GL double layered cloak,  $|\vec r_s| >R_3,$ the excited EM wave field outside of the double layered cloak is as same as in free space and never be disturbed 
by the double layered cloak; (4) provide the source is located outside 
of double layered cloak or located inside of the outer layer of GL cloak, $|\vec r_s| >R_2,$  
the excited EM wave field never propagate into the inner layer of GL cloak and the concealment.

\subsection {There Exists No Maxwell EM Wavefield Can Be Excited By Nonzero
Local Sources Inside Of The Single Layered Cloaked Concealment With Normal Materials}

$\textbf{Statement 2:}$ Suppose that a 3D anisotropic inhomogeneous single layered
cloak domain separates the whole 3D space into three sub domains, 
one is the single layered cloak domain $\Omega _{clk}$ with the cloak material; the second one is the cloaked 
concealment domain $\Omega _{conl}$ with normal EM materials; other one is the 
free space outside of the cloak. If the Maxwell EM wavefield excited 
by a point source or local sources outside of the concealment $\Omega _{conl}$ is vanished
inside of the concealment $\Omega _{conl}$, then there is no Maxwell EM wave field excited 
by the local sources inside of the cloaked concealment $\Omega _{conl}$.

The Maxwell EM wavefield is the EM wave field which satisfies the Maxwell equation
and tangential continuous interface boundary conditions. We call the Maxwell EM wavefield 
as the EM wavefield and use inverse process to prove the $statement \ 2$ as follows:
Suppose that there exists Maxwell EM wavefield excited by the local sources inside the 
concealment with the normal materials, the wavefield satisfies the Maxwell equation in the 
3D whole space $R^3$ which includes the
anisotropic inhomogeneous cloak domain $\Omega _{clk}$  and concealment  $\Omega _{conl}$, and satisfies
the tangential continuous interface conditions on the interface boundary surface $S_1$ and $S_2$. 
The $S_1$ is the interface boundary surface between the
cloak domain $\Omega _{clk}$ and the concealment $\Omega _{conl}$, it also is 
the inner boundary surface of the cloak domain $\Omega _{clk}$.
The $S_2$ is the interface boundary surface between the
cloak domain $\Omega _{clk}$ and the free space, it also is 
the outer boundary surface of the cloak domain $\Omega _{clk}$.

Let $R_{c}  = R^3  - \Omega _{clk} \bigcup {\Omega _{conl}}$,  
$R_{d}  = R^3  - \Omega _{conl}$,
and by the EM integral equation (1), the EM wave field satisfies
\begin{equation}
\begin{array}{l}
 \left[ {\begin{array}{*{20}c}
   {E\left( {r,t} \right)}  \\
   {H\left( {r,t} \right)}  \\
\end{array}} \right] = \left[ {\begin{array}{*{20}c}
   {E_b \left( {r,t} \right)}  \\
   {H_b \left( {r,t} \right)}  \\
\end{array}} \right] + \\ 
 \int\limits_{\Omega _{clk} \bigcup {\Omega _{con l} } } {G_{E,H}^{J,M} 
\left( {r',r,t} \right) * _t \delta \left[ D \right]\left[ {\begin{array}{*{20}c}
   {E_b \left( {r',t} \right)}  \\
   {H_b \left( {r',t} \right)}  \\
\end{array}} \right]dr'}, \\ 
 \end{array}
\end{equation}
where $G_{E,H}^{J,M} \left( {r',r,t} \right)$ is the EM 
Green's tensor, its components
$E^J$, $H^J$, $E^M$, and $H^M \left( {r',r,t} \right)$ are the
EM Green's function on $\Omega _{clk} \bigcup {\Omega _{conl} } \bigcup {R_c} $, 
excited by the 
point  impulse sources outside of the concealment, $r \in R_d$.
By the assumptions, $G_{E,H}^{J,M} \left( {r',r,t} \right)$ exists on
$\Omega _{clk} \bigcup {\Omega _{conl} } \bigcup {R_c }$
and when $r' \in \Omega _{conl}$, $G_{E,H}^{J,M} \left( {r',r,t} \right) = 0$. 
The integral equation (10) becomes to
\begin{equation}
\begin{array}{l}
 \left[ {\begin{array}{*{20}c}
   {E\left( {r,t} \right)}  \\
   {H\left( {r,t} \right)}  \\
\end{array}} \right] = \left[ {\begin{array}{*{20}c}
   {E_b \left( {r,t} \right)}  \\
   {H_b \left( {r,t} \right)}  \\
\end{array}} \right] \\ 
  + \int\limits_{\Omega _{clk} } {G_{E,H}^{J,M} \left( {r',r,t} \right) * _t \delta \left[ D \right]
\left[ {\begin{array}{*{20}c}
   {E_b \left( {r',t} \right)}  \\
   {H_b \left( {r',t} \right)}  \\
\end{array}} \right]dr'.}  \\ 
 \end{array}
\end{equation}

We consider the Maxwell equation in $R_d$, the
virtual source is located $r$, $r \in R_d$ and the point source is located $r_s$,  
$ r_s  \in \Omega _{conl} $ 
and $ r_s  \notin R_d$, we have
\begin{equation}
\begin{array}{l}
 \left[ {\begin{array}{*{20}c}
   {} & {\nabla  \times }  \\
   { - \nabla  \times } & {}  \\
\end{array}} \right]G_{E,H}^{J,M} \left( {r',r,t} \right) \\ 
  = \left[ D \right]G_{E,H}^{J,M} \left( {r',r,t} \right) + I\delta (r',r)\delta (t), \\ 
 \end{array}
\end{equation}
and
\begin{equation}
\begin{array}{l}
 \left[ {\begin{array}{*{20}c}
   {} & {\nabla  \times }  \\
   { - \nabla  \times } & {}  \\
\end{array}} \right]\left[ {\begin{array}{*{20}c}
   {E_b }  \\
   {H_b }  \\
\end{array}} \right]\left( {r',r_s,t} \right) \\ 
  = \left[ {D_b } \right]\left[ {\begin{array}{*{20}c}
   {E_b }  \\
   {H_b }  \\
\end{array}} \right]\left( {r',r_s,t} \right), \\ 
 \end{array}
\end{equation}
By using $ \left[ {E_b \left( {r,t} \right),H_b \left( {r,t} \right)} \right]$ to 
convolute (12), and $G_{E,H}^{J,M} \left( {r',r,t} \right)$ to convolute (13), 
to subtract the second result equation from the first result equation and make 
their integral in ${R_d }$, and use integral by part and make some manipulations, 
we can prove
\begin{equation}
\begin{array}{l}
 \left[ {\begin{array}{*{20}c}
   {E_b \left( {r,t} \right)}  \\
   {H_b \left( {r,t} \right)}  \\
\end{array}} \right] +  \\ 
  + \int\limits_{\Omega _{clk} } {G_{E,H}^{J,M} \left( {r',r,t} \right) * _t \delta \left[ D \right]\left[ {\begin{array}{*{20}c}
   {E_b \left( {r',t} \right)}  \\
   {H_b \left( {r',t} \right)}  \\
\end{array}} \right]dr'}  \\ 
  =  \oint\limits_{S_1 } {G_{E,H}^{J,M} \left( {r',r,t} \right) \otimes_t } \left[ {\begin{array}{*{20}c}
   {E_b \left( {r',t} \right)}  \\
   {H_b \left( {r',t} \right)}  \\
\end{array}} \right]d\vec S, \\ 
 \end{array}
\end{equation}

where $\otimes_t$ is cross convolution.  From the assumption of the $statement \ 2$ that "the Maxwell EM wavefield excited 
by a point source or local sources outside of the concealment $\Omega _{conl}$ is
vanished
in inside of the concealment $\Omega _{conl}$", and virtual source $r$ is located 
outside
of the concealment, $r \in R_d$, if $ r' \in \Omega _{conl} $, 
$G_{E,H}^{J,M} \left( {r',r,t} \right) = 0$. By continuity,    
when $ r' \in S_1 $, we have 
$G_{E,H}^{J,M} \left( {r',r,t} \right) = 0$.  By tangential continuous interface conditions of
$G_{E,H}^{J,M} \left( {r',r,t} \right)$, the term in right hand side of (14) is vanished, we have
\begin{equation}
\begin{array}{l}
 \left[ {\begin{array}{*{20}c}
   {E_b \left( {r,t} \right)}  \\
   {H_b \left( {r,t} \right)}  \\
\end{array}} \right] +  \\ 
  + \int\limits_{\Omega _{clk} } {G_{E,H,b}^{J,M} \left( {r',r,t} \right) * _t \delta \left[ D \right]\left[ {\begin{array}{*{20}c}
   {E\left( {r',t} \right)}  \\
   {H\left( {r',t} \right)}  \\
\end{array}} \right]dr' = 0}. \\ 
 \end{array}
\end{equation}

Upon substituting integral equation (15) into the integral equation (11), we have
\begin{equation}
\left[ {\begin{array}{*{20}c}
   {E\left( {r,t} \right)}  \\
   {H\left( {r,t} \right)}  \\
\end{array}} \right] = 0.
\end{equation}

From the continuous property of the EM wave field, we obtain the following over vanish boundary
condition on the boundary $S_1$ of the concealment $\Omega _{conl}$, we have
\begin{equation}
\left. {\left[ {\begin{array}{*{20}c}
   {E\left( {r,r_s,t} \right)}  \\
   {H\left( {r,r_s,t} \right)}  \\
\end{array}} \right]} \right|_{S_1 }  = 0,
\end{equation}
where $r_s$ denotes point source location inside of the concealment, $r$ is EM field receiver point,
$r \in S_1$. 
Because the EM wave field is excited by local sources inside of the concealment domain $\Omega _{conl}$, 
it satisfies
the following Maxwell equation,
\begin{equation}
\begin{array}{l}
 \left[ {\begin{array}{*{20}c}
   {} & {\nabla  \times }  \\
   { - \nabla  \times } & {}  \\
\end{array}} \right]\left[ {\begin{array}{*{20}c}
   E  \\
   H  \\
\end{array}} \right]\left( {r',r_s,t} \right) \\ 
  = \left[ D_{conl} \right]\left[ {\begin{array}{*{20}c}
   E  \\
   H  \\
\end{array}} \right]\left( {r',r_s,t} \right) + Q(r',r_s,t). \\ 
 \end{array}
\end{equation}

where 
$\left[ D_{conl} \right] = diag\left[ {
   \varepsilon_r\varepsilon_0  \  \mu_r\mu_0 } \right](\partial /\partial t)$
with the normal EM material parameters, $\varepsilon_r\ge 1$ and $\mu_r\ge 1$ are relative EM
parameters, $\varepsilon_0$ is basic permittivity and $\mu_0 $ is basic permeability, 
$r_s  \in \Omega _{conl} $ is
the local source location, $Q(r',r_s,t)$ is the nonzero local source inside of $ \Omega _{conl} $.
Let $G_{E,H,conl}^{J,M} (r',r,t)$ be $Green's$ tensor which satisfies 
\begin{equation}
\begin{array}{l}
 \left[ {\begin{array}{*{20}c}
   0 & {\nabla  \times }  \\
   { - \nabla  \times } & 0  \\
\end{array}} \right]G_{E,H,conl}^{J,M} (r',r,t) \\ 
  = \left[ {D_{conl} } \right]G_{E,H,conl}^{I,M} (r',r,t) \\ 
  + I\delta (r',r)\delta \left( t \right) \\ 
 \end{array}
\end{equation}

By using $ \left[ {E \left( {r,t} \right),H \left( {r,t} \right)} \right]$ to 
convolute (19), and $G_{E,H,conl}^{J,M} \left( {r',r,t} \right)$ to convolute (18), 
to subtract the second result equation from the first result equation and make 
their integral in $\Omega _{conl}$, and use integral by part and make some manipulations,
we have
\begin{equation}
\begin{array}{l}
 \left[ {\begin{array}{*{20}c}
   {E(r,r_s ,t)}  \\
   {H(r,r_s ,t)}  \\
\end{array}} \right] \\ 
  = \int\limits_{\Omega _{conl} } {G_{E,H,conl}^{J,M} } (r',r,t) * _t Q(r',r_s ,t)dr' \\ 
  + \oint\limits_{\partial \Omega _{conl} } {G_{E,H,conl}^{J,M} } (r',r,t) \otimes _t \left[ {\begin{array}{*{20}c}
   {E(r',r_s ,t)}  \\
   {H(r',r_s ,t)}  \\
\end{array}} \right]dr', \\ 
 \end{array}
\end{equation}
$\otimes _t $ denotes the cross convolution, and $\partial \Omega _{conl}=S_1$. Because the over vanished condition (17),
\[
\left. {\left[ {\begin{array}{*{20}c}
   {E\left( {r,r_s,t} \right)}  \\
   {H\left( {r,r_s,t} \right)}  \\
\end{array}} \right]} \right|_{S_1 }  = 0,			
\]
we have
\begin{equation}
\begin{array}{l}
 \left[ {\begin{array}{*{20}c}
   {E(r,r_s ,t)}  \\
   {H(r,r_s ,t)}  \\
\end{array}} \right] =  \\ 
  = \int\limits_{\Omega _{conl} } {G_{E,H,conl}^{J,M} } (r',r,t) * _t Q(r',r_s ,t)dr'. \\ 
 \end{array}
\end{equation}
Because ${G_{E,H,conl}^{J,M} } (r',r,t) \ne 0$ and $Q(r',r_s ,t) \ne 0$, so,
\begin{equation}
\left[ {\begin{array}{*{20}c}
   {E\left( {r,r_s,t} \right)}  \\
   {H\left( {r,r_s,t} \right)}  \\
\end{array}} \right] \ne 0.
\end{equation}
From the continuity of the EM wavefield, the nonzero EM wave field (22) results that 
\begin{equation}
\left. {\left[ {\begin{array}{*{20}c}
   {E\left( {r,r_s,t} \right)}  \\
   {H\left( {r,r_s,t} \right)}  \\
\end{array}} \right]} \right|_{S_1 }  \ne 0.
\end{equation}
The EM wavefield is nonzero on the boundary $S_1$ in (23) is an obvious contradiction with
the same EM wavefield is zero on the boundary $S_1$ in (17).
Therefore, we proved that ${ \bf there  \  exists \  no\ Maxwell\  EM \  wave\  field\ can\ be\ excited \ by \  the}$
${\bf nonzero \ local\   sources\  inside\  of\  the\ single\ layered\ cloaked\ }$ 
${\bf concealment.}$
For more simplicity to derive the nonzero EM wavefield (22) from the integral expression
of the EM wave field (21), let the source is point
impulse current source with polarization direction in $\vec x$, i.e., 
\begin{equation}
Q(r,r_s ,t) = \delta \left( {r - r_s } \right)\delta (t)\vec x.
\end{equation}
Upon substituting (24) and $\varepsilon _r  =1.0 \ and\  \mu _r  = 1.0$ into the (21), we
have 
\begin{equation}
\left[ {\begin{array}{*{20}c}
   {E(r,r_s ,t)}  \\
   {H(r,r_s ,t)}  \\
\end{array}} \right] = \left[ {\begin{array}{*{20}c}
   {E_x^J (r,r_s ,t)}  \\
   {H_x^J (r,r_s ,t)}  \\
\end{array}} \right]
\end{equation}
\begin{equation}
E_x^J (r,r_s ,t) = \left[ {\begin{array}{*{20}c}
   {E_{xx} (r,r_s ,t)}  \\
   {E_{xy} (r,r_s ,t)}  \\
   {E_{xz} (r,r_s ,t)}  \\
\end{array}} \right]
\end{equation}
\begin{equation}
\begin{array}{l}
 E_{xx} (r,r_s ,t) \\ 
  =  - \frac{1}{{8\pi ^2 \varepsilon }}\frac{{\partial ^2 }}{{\partial x^2 }}\frac{{\delta \left( {t - \sqrt {\varepsilon \mu } \left| {r - r_s } \right|} \right)}}{{\left| {r - r_s } \right|}} \\ 
  + \frac{1}{{8\pi ^2 }}\mu \frac{{\partial ^2 }}{{\partial t^2 }}\frac{{\delta \left( {t - \sqrt {\varepsilon \mu } \left| {r - r_s } \right|} \right)}}{{\left| {r - r_s } \right|}} \\ 
 \end{array}
\end{equation}
It is obvious that when $r \in S_1$
\begin{equation}
\left. {E_{xx} (r,r_s ,t)} \right|_{r \in S_1 }  \ne 0.
\end{equation}
The electric intensity $\left. {E_{xx} (r,r_s ,t)} \right|_{r \in S_1 }  \ne 0$ in (28)
and $\left. {E_{xx} (r,r_s ,t)} \right|_{r \in S_1 }  = 0.$ in (17)
are an obvious contradiction. Therefore, we proved the $Statement \ 2$
that there exists no Maxwell EM wavefield can be excited by the nonzero
local sources inside of the single layered cloaked concealment with normal materials.

\section{\label{sec:level1}The GL EM Modeling Simulations of The EM Wave Field 
Through The GL Double Cloaks}
\subsection{The Simulation Model of The GL Double Layered Cloak}
The simulation model:
the 3D domain is $ [-0.5m,0.5 m] \times [-0.5m,0.5 m]  \times [-0.5m,0.5m]$, 
the mesh number is $201 \times 201 \times 201$, the mesh size is 0.005m. 
The electric current point source is defined as
\begin{equation}
\delta (r - r_s )\delta (t)\vec e,
\end{equation}
where the $r_s$ denotes the location of the point source,
the unit vector $\vec e$ is the polarization direction, 
the time step $dt = 0.3333 \times 10^{ - 10}$ 
second, the frequency band is from 0.05 GHz to 15 GHz, the  largest frequency $f=15 GHz$,  
the shortest wave length is $0.02m$.  
The EM GL double layered cloak $\Omega _{GL}  = \Omega _{GLI} \bigcup {\Omega _{GLO} }$ 
is consist of the double spherical annular $\Omega _{GLI}$ and $\Omega _{GLO} $ 
with the center in the origin and interior radius 
$R_1=0.22m$, meddle radius  $R_2=0.3m$. and exterior radius $R_3=0.35m$. The cloak is divided into 
$90 \times 180 \times 90$ cells. 
The spherical coordinate is used in the sphere 
$r \le R_3$, the Cartesian rectangular coordinate is used in outside $\Omega _{GL}$  to mesh the domain. 


\begin{figure}[h]
\centerline{\includegraphics[width=0.86\linewidth,draft=false]{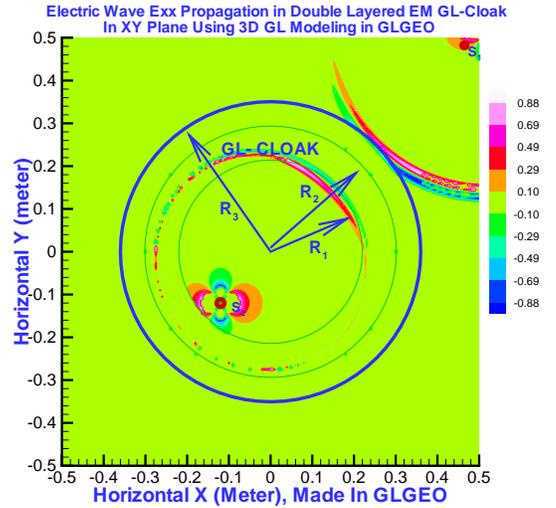}}
\caption{ (color online)  At this moment of the time step 39dt, the most part of the front of the $ {\it First \  electric\  wave} $,  $E_{xx,1}$, propagates enter to the inner GL cloak layer, $R_1 \le r \le R_2$, a few part of the front of the $E_{xx,1}$ is located in right and top of the concealment;  the front of $ {\it Second \  electric\  wave} $,  $E_{xx,2}$,  reaches the outer boundary $r=R_3$ of the
GL double layered cloak.}\label{fig1}
\end{figure}


\begin{figure}[h]
\centerline{\includegraphics[width=0.86\linewidth,draft=false]{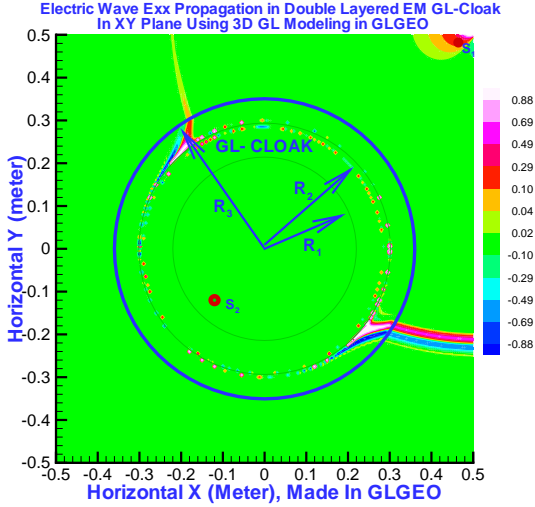}}
\caption{ (color online) At time step $75dt$, the $ {\it Second \  electric\  wave}$, $E_{xx,2}$,
is propagating inside of the outer layer cloak of the GL double layered cloak, 
$R_2 \le r \le R_3$, and around the sphere annular $r=R_2$ and never penetrate into inner domain, 
${r < R_2}$. It does disperse and split into the two phases around the sphere annular $r=R_2$. 
The $ {\it First \  electric\  wave} $,  $E_{xx,1}$,is propagating inside 
of the inner sphere annular layer of the GL double layered cloak, $R_1 \le r \le R_2$.}\label{fig2}
\end{figure}
\begin{figure}[h]
\centerline{\includegraphics[width=0.86\linewidth,draft=false]{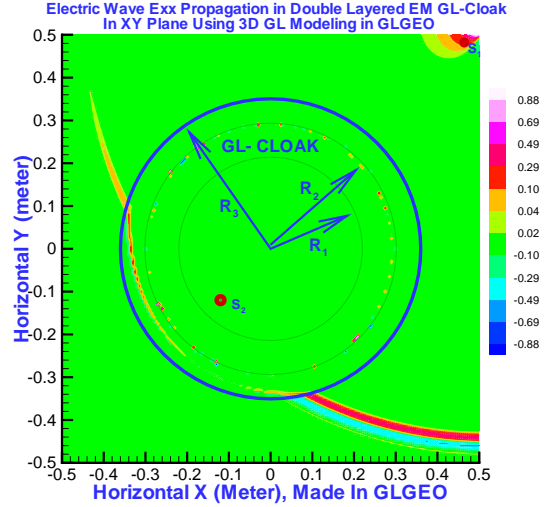}}
\caption{ (color online)  At time step $98dt$, one part of front 
the $ {\it Second \  electric\  wave}$, $E_{xx,2}$,  is propagating inside of the outer layer cloak
, $R_2 \le r \le R_3$. It has 
around the sphere annular $r=R_2$ and forward bending in the left of the sphere annular $r=R_2$. The $ {\it First \  electric\  wave} $,  $E_{xx,1}$,   is still propagating inside 
of the inner sphere annular layer of the GL double layered cloak, $R_1 \le r \le R_2$.}\label{fig2}
\end{figure}

\subsection{The EM Wave Excited By Point Source In The Concealment And Other Point Source
In The Free Space Propagates Through The GL Double Layered Cloak}

The configuration of the GL double layered cloak
material is described in the subsection A of this section. Two point sources are used to excite
the EM wave propagation through the GL double layered cloak.
The first point current source is located inside of the center sphere  concealment 
at $(-0.12m,-0.12m,0.0)$,
by which the
excited EM wave is named as $ {\it First \  EM\  wave} $,  its component $E_{xx,1}$ is named 
$ {\it First \  electric \  wave} $,  
The second current point source is located in free space at $(0.518m,0.518m,0.0)$ where 
is the right and top corner outside of the whole GL double layered cloak. 
The EM wave by the second source is named as $ {\it Second \  EM\  wave} $.
Its component $E_{xx,2}$, is named $ {\it Second \  electric \  wave} $.
The GL modeling simulations of the EM wave
excited by above two point sources
propagation through the GL double layered cloak
are presented in the Figures 1-3. 
The two waves are propagating at time step $38dt$  
that is shown in the Figure 1, at this moment, the most part of the front of the 
$ {\it First \  electric\  wave} $,  $E_{xx,1}$, propagates enter to the 
inner GL cloak layer, $R_1 \le r \le R_2$, a few part of the front of the $E_{xx,1}$
is located in right and top of the concealment.  The front of
$ {\it Second \  electric\  wave} $,  $E_{xx,2}$,  reaches the outer boundary $r=R_3$ of the
GL double layered cloak. In the Figure 2, at time step $75dt$, 
the $ {\it Second \  electric\  wave}$, $E_{xx,2}$,  is propagating inside of the outer layer cloak
of the GL double layered cloak, $R_2 \le r \le R_3$,
and around the sphere annular $r=R_2$ and never penetrate into the inner layer of GL cloak and the concealment
, i.e. $r.le. R_2$. It does disperse and split into the two phases around the sphere annular $r=R_2$ , 
the front phase speed exceeds the light speed; the back phase is slower than the light speed. 
The $ {\it First \  electric\  wave} $,  $E_{xx,1}$,   is propagating inside 
of the inner sphere annular layer of the GL double layered cloak, $R_1 \le r \le R_2$.
In the figure 3, at time step $98dt$, one part of front 
the $ {\it Second \  electric\  wave}$, $E_{xx,2}$,  is propagating inside of the outer layer cloak
, $R_2 \le r \le R_3$. It has 
around the sphere annular $r=R_2$ and forward bending in the left of the sphere annular $r=R_2$  
and never penetrate into the inner layer of GL cloak and the concealment, i.e. $r \le R_2$ 
The $ {\it First \  electric\  wave} $,  $E_{xx,1}$,   is still propagating inside 
of the inner sphere annular layer of the GL double layered cloak, $R_1 \le r \le R_2$.
At time step $138dt$, 
the $ {\it Second \  electric \  wave}$,  $E_{xx,2}$
has propagated outside of the whole GL double layered cloak, a small part of its wave front is 
located in the left and
low corner of the plot frame which is shown in the
Figure which is omitted, most part of front of the $E_{xx,2}$ electric wave field has been out
of the plot frame. The exterior EM wave outside  of the GL double layered cloak 
never been disturbed 
by the cloak and never penetrate enter
the centre sphere concealment and inner layer of the GL cloak. 
At same time step, the $ {\it First \  electric \  wave}$,  $E_{xx,1}$, 
is propagating inside of the 
the inner layer sphere annular of the GL double cloak, $R_1 \le r \le R_2$. It can be very 
closed to the interface boundary $r = R_2$, However, it can not be reached to 
the interface boundary $r = R_2 $ for any long time.


\begin{figure}[h]
\centerline{\includegraphics[width=0.86\linewidth,draft=false]{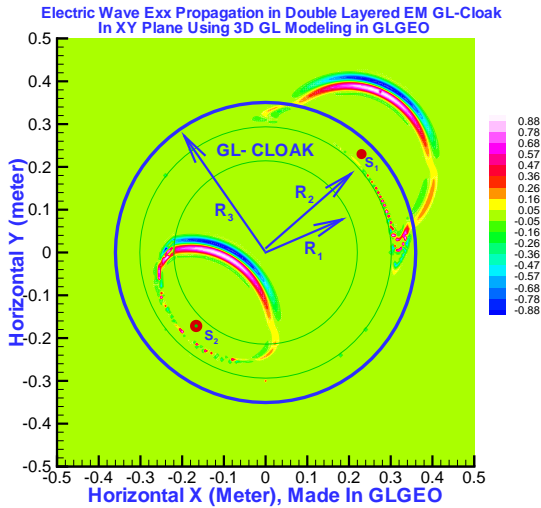}}
\caption{ (color online)
At moment $21dt$,  
one part of the front of the 
$ {\it First \  electric\  wave} $,  $E_{xx,1}$, propagates enter to the 
concealment; other part of the front is still propagating inside of
inner GL cloak layer, $R_1 \le r \le R_2$.  One part of front of
$ {\it Second \  electric\  wave} $,  $E_{xx,2}$,  reaches the middle interface boundary $r=R_2$ of the
GL cloak; other part of the front has propagated in free space with disturbance.}\label{fig4}
\end{figure}

\subsection{The EM Wave Excited By The Point Source In The Inner Layer $\Omega _{GLI}$  And Other Point Source
In The Outer Layer $ \Omega _{GLO} $ Propagates Through The GL Double Layered Cloak}
The 3D EM full wave excited by the point source in the inner layer and other point source in the outer layer of GL double
Layered cloak are simulated by using GL EM modeling. The simulations of the EM wave excited by the above sources through GL double layered cloak are
presented in the Figures 4. The configuration of the GL double layered cloak
material is described in the subsection A of this section. Two point sources are used to excite
the EM wave propagation through the GL double layered cloak.
The first point current source is located inside of the inner layer of the GL cloak at $(-0.165m,-0.165m,0.0)$, by which the
excited EM wave is named as $ {\it First \  EM\  wave} $.  Its component $E_{xx,1}$ is named $ {\it First \  electric \  wave} $.  
The second current point source is located in outer layer GL cloak  at $(0.23m,0.23m,0.0)$. 
The EM wave by the second source is named as $ {\it Second \  EM\  wave} $.
Its component $E_{xx,2}$, is named $ {\it Second \  electric \  wave} $.
In the Figure 4, at moment $21dt$,  
one part of the front of the 
$ {\it First \  electric\  wave} $,  $E_{xx,1}$, propagates enter to the 
concealment; other part of the front is still propagating inside of
inner GL cloak layer, $R_1 \le r \le R_2$.  One part of front of
$ {\it Second \  electric\  wave} $,  $E_{xx,2}$,  reaches the middle interface boundary $r=R_2$ of the
GL cloak; other part of the front has propagated outside of the whole GL double
layered cloak and in free space with disturbance.  At the time step $48dt$, the 
${\it First \  electric\  wave} $, 
$E_{xx,1}$, has  propagated through the concealment and whole front is inside of the inner layer of the GL cloak
$R_1 \le r \le R_2$, its speed is smaller than the light speed. The $ {\it Second \  electric\  wave} $ never propagates into the inner layer of GL cloak,   The part of the front of
 $ {\it Second \  electric\  wave}$,  $E_{xx,2}$, is inside of
outer layer of the GL double layered cloak, 
$R_2 \le r \le R_3$, and being forward bending with speed lager than the light speed. Other part of the front has been propagating in free space 
with disturbance.
The EM wave propagation image is presented in an omitted Figure.  In the other omitted Figure, at time step $68dt$, 
the $ {\it Second \  electric\  wave}$, $E_{xx,2}$,  is propagating in free space and
 outside of the whole GL double layered cloak,  and never penetrate into inner domain, 
${r < R_2}$, i.e. never penetrate into 
the inner layer of GL cloak and the concealment. 
The $ {\it First \  electric\  wave} $,  $E_{xx,1}$,   is still propagating inside 
of the inner sphere annular layer of the GL double layered cloak, $R_1 \le r \le R_2$.
At time step $98dt$, 
the $ {\it Second \  electric\  wave}$, $E_{xx,2}$,  has been propagating out of 
whole GL cloak and out of the plot frame.
Very small part of its front is located in
left and low corner of the plot frame, the fihure is omitted.
The $ {\it First \  electric\  wave} $,  $E_{xx,1}$,   is still propagating inside 
of the inner sphere annular layer of the GL double layered cloak, $R_1 \le r \le R_2$.
It can be very 
closed to the interface boundary $r = R_2$, However, it can not be reached to 
the interface boundary $r = R_2 $ for any long time.

\section{\label{sec:level1}Advantages}
\subsection {The EM GL Double Layered Cloak Is Robust For Invisibility}
The figure 4 clearly show that the 
wave front of the Second electric wave has propagated outside of the 
cloak and go to free space with disturbance. The results reminders us that 
if only single outer layer cloak $\Omega_{GLO}$ is adopted, and there is 
a little crack loss on the inner boundary surface $\partial{{\Omega_{GLO}}_-}$, 
some EM or current source inside of the $\Omega_{GLO}$
will excite the EM wave propagation go out to free space and expose the 
cloak immediately. The GL double layered cloak overcomes the weakness that
is also shown in the figure 4.
The wave front of the First electric wave, which is excited by a point
source inside of the inner layer cloak $\Omega_{GLI}$ , is always propagating 
inside of inner layer cloak $\Omega_{GLI}$ or
concealment $\Omega_{conl}$ and never propagate outside of the interface 
annular $ S_1 $. Therefore, the EM GL double layered cloak is robust
for invisibility.

\subsection {The EM GL Double Layered Cloak Is Complete Invisible}
The figure 1-3 clearly show that the outer layer cloak of the GL double layered 
cloak has the invisibility function,
while the inner layer cloak has fully absorption function to absorb
the EM wave excited from local sources inside of the 
concealment.
When the  source is located outside 
of the GL double layered cloak, the excited EM wave field 
propagation outside of the double layered cloak is as same as in free space and never be 
disturbed 
by the cloak; also, the exterior EM wave can not penetrate into the
inner layer and concealment of the GL double layered cloak.
When local sources are located inside of the GL double cloaked concealment
with the normal EM materials, the 
excited EM wave is propagating under Maxwell equation governing,
it is complete and successively absorbed by the inner layer cloak of GL double cloak 
and never propagate to outside of the inner layer of the GL cloak,
moreover, the EM wavefield in concealment never be disturbed by the cloak. 

Using the GL method theoretical analysis, the statement 2 in section 4 is rigorous 
proved. It states that "there exists no Maxwell electromagnetic wavefield can be
excited by nonzero local sources inside of the single layer cloaked concealment
with the normal EM materials". The invisibility of the single layered cloak
and existence of Maxwell EM wave field excited by the local sources inside 
its concealment is inconsistent. Provide only single outer layered cloak
is adopted. The EM field excited by local sources inside of its concealment with
normal materials does not satisfy the Maxwell equation. The EM chaos phenomena, 
which is divorced from the Maxwell equation governing, may damage devices
and human inside the concealment, or may degrade the invisibility of the cloak.
Invisibility function of the single layered cloak is not complete.
The EM GL double layered cloak overcomes the drawback of the single layered cloak. The GL double layered cloak has the complete sufficient invisibility function.

\subsection {The EM GL Double Layered Cloak Can Be Double Ellipsoid Annular}
The EM GL double layered cloak can be extended to have double ellipsoid annular
and other geometrical double layered closed strips. 

\subsection {Frequency Band}

Many simulations and theoretical analysis by the GL method show that the idea
EM GL double layered cloak is of the invisibility function for all frequencies. 
However, the practical material has some loss.
The frequency band will be depended on the rate of the material loss. 
Because the EM GL double layered cloak is robust and complete cloak,
it has three radius $R_1$, $R_2$, and $R_3$ can be chosen, and
GL method reduced numerical frequency limitation in FEM and FD scheme,
a reasonable wide frequency band of the GL double layered cloak for low loss rate 
will be optimization and reported in next paper.

\subsection {Advantages Of The GL Method}
The GL EM modeling is fully different from FEM and FD and Born approximation methods and overcome their difficulties. There is no big matrix equation to solve in GL method.
Moreover, it does not need artificial boundary and absorption condition
to truncate the infinite domain. 
Born Approximation is a conventional method in the quantum mechanics
and solid physics, however, it is one iteration only in whole domain which is
not accurate for high frequency and for high contrast materials. The GL method divides the domain as a set of small 
sub domains or sub lattices. The Global field is updated by the local field 
from the interaction between the global field  
and local subdomain materials successively. Once all subdomain
materials are scattered, the GL field solution is obtained which
is much more accurate than the Born approximation. GL method
is suitable for all frequency and high contrast materials.
 
Moreover, the GL method can be meshless, including arbitrary geometry
subdomains, such as rectangle, cylindrical and spherical coordinate
mixed coupled together. It is full parallel algorithm.  
These advantages of the GL method to overcome historical
difficulties have been detailed described in the paper [1]. 
The theoretical foundation  of the GL method is described in the paper [2]
The GL EM method consistent combines the 
analytical and numerical approaches together and reduced the numerical 
dispersion and numerical frequency limitation. The GL method has double abilities of
the theoretical analysis and numerical simulations that
has been shown in this paper.

The 3D GL simulations of the EM wave field through the single and multiple sphere, cylinder, ellipsoid, and arbitrary geometry cloaks in single layered and double layered profile
show that the GLT and GLF EM modeling are accurate, stable and fast.
It saves much more storages than the conventional methods. In general, only 10 to 50 minute are needed to run the 3D EM wave field through the cloaks with 64 to 128 frequencies in the PC. The high performance
GL parallel algorithm in PC cluster
and super parallel computer is very fast and powerful to simulate
complex and large scale physical and chemical process.

A double layer cloth phenomenon to prevent the GILD inversion [5-7]
detection has been observed in paper [8] in 2001 which is published in SEG online
http://www.segdl.org/journals/doc/SEGLIB-home/dci/searchDCI.jsp. 
After the event, we effort improve GILD [5-7] and developed a novel and effective Global and Local field (GL) modeling and inversion to
study the meta materials, periodic photonic crystals and condense
physics etc. wide physical sciences.
3D GL EM modeling and inversion [9] and computational mirage [10] have been presented in PIERS 2005 
and published in
proceeding of PIERS 2005 in Hangzhou, which can be downloaded from 
http://piers.mit.edu/piersproceedings/piers2k5Proc.php.
We developed 3D FEM for the elastic mechanics first in China in 1972 which has been cited and recorded in [17]. Our 3D FEM paper has been
published in [18] in Chinese.
We are deeply and clearly to know the merits and drawbacks of FEM and its serious limitation and difficulties to simulate high frequency wave
propagation through dispersive materials. The GL method
overcome the drawbacks of FEM and FD methods.
The history of development of our 3D FEM [18], novel inversion [7], GILD [5] and GL method [1-3] has been described in [2].
The 3D and 2D GL parallel software is made and patented by GLGEO.
The GL modeling and its inversion [1-3] and GL EM quantum field modeling are suitable
to solve quantization scattering problem of the electromagnetic field in the dispersive and loss metamaterials, cloaks
and more wide anisotropic condense materials.

\section{\label{sec:level1}CONCLUSIONS}

Many simulations of the EM wave propagation through the GL double
layered cloak by the GL modeling and theoretical analysis verify that the EM GL doubled
cloak is robust cloak and has complete and sufficient
invisibility functions. Its concealment is the normal electromagnetic
environment.
The outer layer cloak of the GL double layered cloak has the invisible function,
the inner layer cloak has fully absorption function.
When the source is located outside 
of the GL double layered cloak, the excited EM wave field 
propagation outside of the double layered cloak is as same as in free space and never be 
disturbed 
by the cloak; also, the exterior EM wave can not penetrate into the
inner layer and concealment.
When sources are located inside of the GL double cloaked concealment
with the normal EM materials, the 
excited EM wave is propagating under Maxwell equation governing,
it is complete absorbed by the inner layer cloak of GL double cloak 
and never propagate to outside of the inner layer of the GL cloak,
moreover, the EM wavefield in concealment never be disturbed by the cloak. 
The EM GL double layered cloak has advantages to overcome
the drawback and difficulty of the single layered cloak.

The GL method is an effective physical simulation method.
It has double abilities of the theoretical analysis and numerical simulations to study the cloak metamaterials and wide material and 
Field scattering in physical sciences.

\begin{acknowledgments}
We wish to acknowledge the support of the GL Geophysical Laboratory.
Authors thank to Professor P. D. Lax for his concern and encouragements.
\end{acknowledgments}


\end{document}